\documentclass[aps,prl,twocolumn,showpacs,superscriptaddress,nofootinbib]{revtex4-1}

\usepackage{amsmath}   
\usepackage{graphicx}  
\usepackage{xspace}
\usepackage{units}
\usepackage{accents}
\usepackage{scalerel}
\usepackage{hyperref}

\newlength{\heightnu}
\AtBeginDocument{\settoheight{\heightnu}{$\nu$}}
\newcommand{\nupbar}{\llap{\raisebox{\heightnu+0.5pt}{\scaleobj{0.3}{\hstretch{1.3}(}}}  \overline{\nu}\rlap{\raisebox{\heightnu+0.5pt}{\scaleobj{0.3}{\hstretch{1.3})}}}}

\newcommand{\sizecheck}{0} 
\newcommand{\PRLsupp}{1}   
\ifnum\PRLsupp=0
  
\else
  
\fi

\newif\ifpdf
\ifx\pdfoutput\undefined
   \pdffalse
\else
   \pdfoutput=1
   \pdftrue
\fi
\ifpdf
   \usepackage{graphicx}
   \usepackage{epstopdf}
\else
   \usepackage{graphicx}
\fi

\begin{document}

\title{
Search for CP violation in Neutrino and Antineutrino Oscillations by the T2K experiment with $2.2\times10^{21}$ protons on target\\
}


\newcommand{\INSTHD}{\affiliation{University Autonoma Madrid, Department of Theoretical Physics, 28049 Madrid, Spain}}
\newcommand{\INSTEE}{\affiliation{University of Bern, Albert Einstein Center for Fundamental Physics, Laboratory for High Energy Physics (LHEP), Bern, Switzerland}}
\newcommand{\INSTFE}{\affiliation{Boston University, Department of Physics, Boston, Massachusetts, U.S.A.}}
\newcommand{\INSTD}{\affiliation{University of British Columbia, Department of Physics and Astronomy, Vancouver, British Columbia, Canada}}
\newcommand{\INSTGA}{\affiliation{University of California, Irvine, Department of Physics and Astronomy, Irvine, California, U.S.A.}}
\newcommand{\INSTI}{\affiliation{IRFU, CEA Saclay, Gif-sur-Yvette, France}}
\newcommand{\INSTGB}{\affiliation{University of Colorado at Boulder, Department of Physics, Boulder, Colorado, U.S.A.}}
\newcommand{\INSTFG}{\affiliation{Colorado State University, Department of Physics, Fort Collins, Colorado, U.S.A.}}
\newcommand{\INSTFH}{\affiliation{Duke University, Department of Physics, Durham, North Carolina, U.S.A.}}
\newcommand{\INSTBA}{\affiliation{Ecole Polytechnique, IN2P3-CNRS, Laboratoire Leprince-Ringuet, Palaiseau, France }}
\newcommand{\INSTEF}{\affiliation{ETH Zurich, Institute for Particle Physics, Zurich, Switzerland}}
\newcommand{\INSTEG}{\affiliation{University of Geneva, Section de Physique, DPNC, Geneva, Switzerland}}
\newcommand{\INSTDG}{\affiliation{H. Niewodniczanski Institute of Nuclear Physics PAN, Cracow, Poland}}
\newcommand{\INSTCB}{\affiliation{High Energy Accelerator Research Organization (KEK), Tsukuba, Ibaraki, Japan}}
\newcommand{\INSTED}{\affiliation{Institut de Fisica d'Altes Energies (IFAE), The Barcelona Institute of Science and Technology, Campus UAB, Bellaterra (Barcelona) Spain}}
\newcommand{\INSTEC}{\affiliation{IFIC (CSIC \& University of Valencia), Valencia, Spain}}
\newcommand{\INSTEI}{\affiliation{Imperial College London, Department of Physics, London, United Kingdom}}
\newcommand{\INSTGF}{\affiliation{INFN Sezione di Bari and Universit\`a e Politecnico di Bari, Dipartimento Interuniversitario di Fisica, Bari, Italy}}
\newcommand{\INSTBE}{\affiliation{INFN Sezione di Napoli and Universit\`a di Napoli, Dipartimento di Fisica, Napoli, Italy}}
\newcommand{\INSTBF}{\affiliation{INFN Sezione di Padova and Universit\`a di Padova, Dipartimento di Fisica, Padova, Italy}}
\newcommand{\INSTBD}{\affiliation{INFN Sezione di Roma and Universit\`a di Roma ``La Sapienza'', Roma, Italy}}
\newcommand{\INSTEB}{\affiliation{Institute for Nuclear Research of the Russian Academy of Sciences, Moscow, Russia}}
\newcommand{\INSTHA}{\affiliation{Kavli Institute for the Physics and Mathematics of the Universe (WPI), The University of Tokyo Institutes for Advanced Study, University of Tokyo, Kashiwa, Chiba, Japan}}
\newcommand{\INSTCC}{\affiliation{Kobe University, Kobe, Japan}}
\newcommand{\INSTCD}{\affiliation{Kyoto University, Department of Physics, Kyoto, Japan}}
\newcommand{\INSTEJ}{\affiliation{Lancaster University, Physics Department, Lancaster, United Kingdom}}
\newcommand{\INSTFC}{\affiliation{University of Liverpool, Department of Physics, Liverpool, United Kingdom}}
\newcommand{\INSTFI}{\affiliation{Louisiana State University, Department of Physics and Astronomy, Baton Rouge, Louisiana, U.S.A.}}
\newcommand{\INSTHB}{\affiliation{Michigan State University, Department of Physics and Astronomy,  East Lansing, Michigan, U.S.A.}}
\newcommand{\INSTCE}{\affiliation{Miyagi University of Education, Department of Physics, Sendai, Japan}}
\newcommand{\INSTDF}{\affiliation{National Centre for Nuclear Research, Warsaw, Poland}}
\newcommand{\INSTFJ}{\affiliation{State University of New York at Stony Brook, Department of Physics and Astronomy, Stony Brook, New York, U.S.A.}}
\newcommand{\INSTGJ}{\affiliation{Okayama University, Department of Physics, Okayama, Japan}}
\newcommand{\INSTCF}{\affiliation{Osaka City University, Department of Physics, Osaka, Japan}}
\newcommand{\INSTGG}{\affiliation{Oxford University, Department of Physics, Oxford, United Kingdom}}
\newcommand{\INSTBB}{\affiliation{UPMC, Universit\'e Paris Diderot, CNRS/IN2P3, Laboratoire de Physique Nucl\'eaire et de Hautes Energies (LPNHE), Paris, France}}
\newcommand{\INSTGC}{\affiliation{University of Pittsburgh, Department of Physics and Astronomy, Pittsburgh, Pennsylvania, U.S.A.}}
\newcommand{\INSTFA}{\affiliation{Queen Mary University of London, School of Physics and Astronomy, London, United Kingdom}}
\newcommand{\INSTE}{\affiliation{University of Regina, Department of Physics, Regina, Saskatchewan, Canada}}
\newcommand{\INSTGD}{\affiliation{University of Rochester, Department of Physics and Astronomy, Rochester, New York, U.S.A.}}
\newcommand{\INSTHC}{\affiliation{Royal Holloway University of London, Department of Physics, Egham, Surrey, United Kingdom}}
\newcommand{\INSTBC}{\affiliation{RWTH Aachen University, III. Physikalisches Institut, Aachen, Germany}}
\newcommand{\INSTFB}{\affiliation{University of Sheffield, Department of Physics and Astronomy, Sheffield, United Kingdom}}
\newcommand{\INSTDI}{\affiliation{University of Silesia, Institute of Physics, Katowice, Poland}}
\newcommand{\INSTIA}{\affiliation{SLAC National Accelerator Laboratory, Stanford University, Menlo Park, California, USA}}
\newcommand{\INSTEH}{\affiliation{STFC, Rutherford Appleton Laboratory, Harwell Oxford,  and  Daresbury Laboratory, Warrington, United Kingdom}}
\newcommand{\INSTCH}{\affiliation{University of Tokyo, Department of Physics, Tokyo, Japan}}
\newcommand{\INSTBJ}{\affiliation{University of Tokyo, Institute for Cosmic Ray Research, Kamioka Observatory, Kamioka, Japan}}
\newcommand{\INSTCG}{\affiliation{University of Tokyo, Institute for Cosmic Ray Research, Research Center for Cosmic Neutrinos, Kashiwa, Japan}}
\newcommand{\INSTHF}{\affiliation{Tokyo Institute of Technology, Department of Physics, Tokyo, Japan}}
\newcommand{\INSTGI}{\affiliation{Tokyo Metropolitan University, Department of Physics, Tokyo, Japan}}
\newcommand{\INSTHG}{\affiliation{Tokyo University of Science, Faculty of Science and Technology, Department of Physics, Noda, Chiba, Japan}}
\newcommand{\INSTF}{\affiliation{University of Toronto, Department of Physics, Toronto, Ontario, Canada}}
\newcommand{\INSTB}{\affiliation{TRIUMF, Vancouver, British Columbia, Canada}}
\newcommand{\INSTG}{\affiliation{University of Victoria, Department of Physics and Astronomy, Victoria, British Columbia, Canada}}
\newcommand{\INSTDJ}{\affiliation{University of Warsaw, Faculty of Physics, Warsaw, Poland}}
\newcommand{\INSTDH}{\affiliation{Warsaw University of Technology, Institute of Radioelectronics, Warsaw, Poland}}
\newcommand{\INSTFD}{\affiliation{University of Warwick, Department of Physics, Coventry, United Kingdom}}
\newcommand{\INSTGH}{\affiliation{University of Winnipeg, Department of Physics, Winnipeg, Manitoba, Canada}}
\newcommand{\INSTEA}{\affiliation{Wroclaw University, Faculty of Physics and Astronomy, Wroclaw, Poland}}
\newcommand{\INSTHE}{\affiliation{Yokohama National University, Faculty of Engineering, Yokohama, Japan}}
\newcommand{\INSTH}{\affiliation{York University, Department of Physics and Astronomy, Toronto, Ontario, Canada}}

\INSTHD
\INSTEE
\INSTFE
\INSTD
\INSTGA
\INSTI
\INSTGB
\INSTFG
\INSTFH
\INSTBA
\INSTEF
\INSTEG
\INSTDG
\INSTCB
\INSTED
\INSTEC
\INSTEI
\INSTGF
\INSTBE
\INSTBF
\INSTBD
\INSTEB
\INSTHA
\INSTCC
\INSTCD
\INSTEJ
\INSTFC
\INSTFI
\INSTHB
\INSTCE
\INSTDF
\INSTFJ
\INSTGJ
\INSTCF
\INSTGG
\INSTBB
\INSTGC
\INSTFA
\INSTE
\INSTGD
\INSTHC
\INSTBC
\INSTFB
\INSTDI
\INSTIA
\INSTEH
\INSTCH
\INSTBJ
\INSTCG
\INSTHF
\INSTGI
\INSTHG
\INSTF
\INSTB
\INSTG
\INSTDJ
\INSTDH
\INSTFD
\INSTGH
\INSTEA
\INSTHE
\INSTH

\author{K.\,Abe}\INSTBJ
\author{R.\,Akutsu}\INSTCG
\author{A.\,Ali}\INSTBF
\author{J.\,Amey}\INSTEI
\author{C.\,Andreopoulos}\INSTEH\INSTFC
\author{L.\,Anthony}\INSTFC
\author{M.\,Antonova}\INSTEC
\author{S.\,Aoki}\INSTCC
\author{A.\,Ariga}\INSTEE
\author{Y.\,Ashida}\INSTCD
% ...

\collaboration{The T2K Collaboration}\noaffiliation
\date{\today}

\begin{abstract}
The T2K experiment measures muon neutrino disappearance and electron neutrino appearance in accelerator-produced neutrino and antineutrino beams.
With an exposure of $14.7(7.6)\times 10^{20}$ protons on target in neutrino (antineutrino) mode, 89 $\nu_e$ candidates and 7 anti-$\nu_e$ candidates were observed while 67.5 and 9.0 are expected for $\delta_{CP}=0$ and normal mass ordering.
The obtained $2\sigma$ confidence interval for the $CP$ violating phase, $\delta_{CP}$, does not include the $CP$-conserving cases ($\delta_{CP}=0,\pi$).
The best-fit values of other parameters are $\sin^2\theta_{23} = 0.526^{+0.032}_{-0.036}$ and $\Delta m^2_{32}=2.463^{+0.071}_{-0.070}\times10^{-3} \mathrm{eV}^2/c^4$.
\end{abstract}

\ifnum\sizecheck=0  
\maketitle
\fi




\textit{Introduction. --- }
Observation of neutrino oscillations has established that each of the three flavor states of neutrinos is a superposition of at least three mass eigenstates, $m_1, m_2$ and $m_3$~\cite{Fukuda:1998mi,Ahmad:2001an,Abe:2011sj,An:2012eh}.
As a consequence of three-generation mixing, the flavor-mass mixing matrix, the PMNS matrix~\cite{Maki:1962mu,Pontecorvo:1967fh}, can have an irreducible imaginary component, and $CP$ symmetry can be violated in neutrino oscillations, analogous to the case of the quark sector.
The PMNS matrix is parametrized by three mixing angles $\theta_{12}$, $\theta_{13}$ and $\theta_{23}$, and one $CP$ violation phase, $\delta_{CP}$, which gives rise to asymmetries between neutrino oscillations and antineutrino oscillations if $\sin \delta_{CP} \ne 0$.
The magnitude of $CP$ violation is determined by the invariant $J_{CP}=\frac{1}{8}\cos{\theta_{13}}\sin{2\theta_{12}}\sin{2\theta_{23}}\sin{2\theta_{13}}\sin{\delta_{CP}}\approx 0.033\sin{\delta_{CP}}~$\cite{Krastev:1988yu,Jarlskog:1985cw} and could be large compared to the quark sector value ($J_{CP}\approx 3\times 10^{-5}$).
The most feasible way to probe $\delta_{CP}$ is by measuring the appearance of electron (anti)neutrinos ($\nupbar_{e}$) by using accelerator-produced muon (anti)neutrino ($\nupbar_{\mu}$) beams.
T2K has reported that the $CP$ conservation hypothesis ($\delta_{CP}=0,\pi$) is excluded at 90\% confidence level (C.L.) using the data collected up to May 2016~\cite{Abe:2017uxa,Abe:2017vifw}.
Since then, the neutrino mode data set has doubled, and the electron neutrino and antineutrino event selection efficiencies have increased by 30\% and 20\%, respectively.
In this Letter, we report new results on $\delta_{CP}$, $\sin^2\theta_{23}$ and $\Delta m^2$($\Delta m^2_{32}\equiv m_3^2-m_2^2$ for normal or $\Delta m^2_{13}\equiv m_1^2-m_3^2$ for inverted mass ordering) obtained by analyzing both muon (anti)neutrino disappearance and electron (anti)neutrino appearance data collected up to May 2017 using a new event selection method.

\textit{The T2K experiment}~\cite{Abe:2011ks}\textit{. --- }
The 30 GeV proton beam from the J-PARC accelerator strikes a graphite target to produce charged pions and kaons which are focused or defocused by a system of three magnetic horns. The focused charge is defined by the horn current direction, producing either a muon neutrino or antineutrino beam from the focused secondaries decaying in the 96 m long decay volume.
An on-axis near detector (INGRID) and a detector 2.5$^\circ$ off the beam axis (ND280) sample the unoscillated neutrino beam 280~m downstream from the target station and monitor the beam direction, composition, and intensity.
The off-axis energy spectrum peaks at 0.6 GeV and has significantly less $\nupbar_{e}$ contamination at the peak energy and less high energy neutrino flux than on axis.
The Super-Kamiokande (SK) 50 kt water-Cherenkov detector~\cite{Fukuda:2002uc}, as a far detector, samples the oscillated neutrino beam 2.5$^{\circ}$ off axis and 295 km from the production point.

\textit{Data set. --- }
The results presented here are based on data collected from Jan 2010 to May 2017.
The data sets include a beam exposure of $14.7\times 10^{20}$ protons on target (POT) in neutrino mode and $7.6\times 10^{20}$ POT in antineutrino mode for the far-detector (SK) analysis, and an exposure of 5.8$\times 10^{20}$ POT in neutrino mode and 3.9$\times 10^{20}$ POT in antineutrino mode for the near-detector (ND280) analysis.

\textit{Analysis strategy. --- }
Oscillation parameters are determined by comparing model predictions with observations at the near and far detectors.
The neutrino flux is modeled based on a data-driven simulation.
The neutrino-nucleus interactions are simulated based on theoretical models with uncertainties estimated from data and models.
The flux and interaction models are refined by the observation of the rate and spectrum of charged-current (CC) neutrino interactions by ND280. 
Since ND280 is magnetized, wrong-sign contamination in the beam can be estimated from charge-selected near-detector samples.
The prediction of the refined model is compared with the observation at SK to estimate the oscillation parameters.
Overall analysis method is the same as in previous T2K results~\cite{Abe:2017vifw}, but
this analysis uses improved theoretical models to describe neutrino interactions and a new reconstruction algorithm at SK, which improves signal/background discrimination and allows an expanded fiducial volume.

\textit{Neutrino flux model. --- }
A data-driven simulation is used to calculate the neutrino and antineutrino fluxes and their uncertainties at each detector, including correlations~\cite{Abe:2017vifw,Abe:2013fp}.
The interactions of hadrons in the target and other beamline materials are tuned using external thin-target hadron-production data, mainly measurements of 30~GeV protons on a graphite target by the NA61/SHINE experiment~\cite{Abgrall:2015hmv}.
The simulation reflects the proton beam condition, horn current and neutrino beam-axis direction as measured by monitors.
Near the peak energy, and in the absence of oscillations, 97.2\%\ (96.2\%) of the (anti)neutrino mode beam is $\nupbar_{\mu}$. 
The remaining components are mostly $\bar{\nu}_\mu$($\nu_\mu$); contamination of $\nupbar_{e}$ is only 0.42\%\ (0.46\%).
The dominant source of systematic error in the flux model is the uncertainty of the hadron-production data.
Some of the beamline conditions are different depending on time.
Stability of the neutrino flux has been monitored by INGRID throughout the whole data taking period.
The flux covariance matrix was constructed by removing the near-far correlations for time-dependent systematics for the period during which ND280 data was not used in this analysis.
While the flux uncertainty is approximately 9\% at the peak energy, its impact on oscillation parameter uncertainties, given that the near and far detector measurements sample nearly the same flux, is significantly smaller.

{\it Neutrino interaction model. --- } Events are simulated with the NEUT~\cite{NEUT} neutrino interaction generator. The dominant charged-current quasi-elastic (CCQE)-like interaction (defined as those with a charged lepton, and no pions in the final state) is modeled with a relativistic Fermi gas (RFG) nuclear model including long-range correlations using the random phase approximation (RPA)~\cite{nieves_rpa,*nieves_rpa_erratum}. The 2p-2h model of Nieves {\em et al.}~\cite{nieves, nievesExtension} predicts multi-nucleon contributions to CCQE-like processes
.
These can be divided into meson exchange current (“$\Delta$-like”) contributions, which include both diagrams with an intermediate $\Delta$ and contributions from pion in-flight and pion contact terms (see Ref.~\cite{nieves} for details), and contributions from interactions with correlated nucleon–nucleon pairs (“non-$\Delta$-like”), which introduce different biases in the reconstructed neutrino energy, $E_\mathrm{rec}$, calculated assuming QE scattering~\cite{Abe:2017vifw}\footnote{Figure 5 of \cite{Abe:2017vifw} shows the quantitative difference.}.
New parameters are introduced to vary the relative contribution of $\Delta$-like
and non-$\Delta$-like terms\footnote{There is an interference term between the two terms which is rescaled to preserve the total 2p-2h cross section, but is not recalculated.} for $^{12}$C and $^{16}$O, with a 30\% correlation between the two nuclei.
The total 2p-2h normalization is varied separately for $\nu$ and $\bar{\nu}$ with flat priors. There is an additional uncertainty on the ratio of $^{12}$C to $^{16}$O 2p-2h normalizations, with a 20\% uncertainty. The $Q^{2}$ dependence of the RPA correction is allowed to vary by the addition of four variable parameters designed to span the total theoretical uncertainty in the $Q^{2}$ dependence~\cite{nieves_rpa_uncertainty,rik_rpa}. Processes producing a single pion and one or more nucleons in the final state are described by the Rein-Sehgal model~\cite{ReinSehgal}. Parameters describing the $\Delta$ axial form factor and single pion production not through baryon resonances are tuned to match $D_2$ measurements~\cite{ANL_Radecky_1982, BNL_Kitagaki_1986, D2-ANLBNL-fix} in a method similar to \cite{Wilkinson-Rodrigues-GENIEtune}. Production of pions in coherent inelastic scattering is described by a tuned model of Rein-Sehgal~\cite{RS-coherent} which agrees with recent measurements~\cite{T2K-coherent,MINERvA-coherent}.
As in \cite{Abe:2017vifw}, differences between muon- and electron-neutrino interactions occur because of final-state lepton mass and radiative corrections, and are largest at low energies.
To account for this, we add a 2\% uncorrelated uncertainty for each of 
electron neutrino and antineutrino cross sections relative to those of muons ($\sigma^{\mathrm{CC}}(\nu_{e})/\sigma^{\mathrm{CC}}(\nu_{\mu})$ and $\sigma^{\mathrm{CC}}(\bar{\nu}_{e})/\sigma^{\mathrm{CC}}(\bar{\nu}_{\mu})$), and another 2\% uncertainty anticorrelated between the two ratios~\cite{McFarland-Day}.
The cross-section parametrization is otherwise as described in \cite{Abe:2017vifw}, with the exception of variations of the nucleon removal energy, $E_{\mathrm{b}}$, by 25(27)$^{+18}_{-9}$ MeV for $^{12}$C($^{16}$O)~\cite{Bodek:2018lmc}.

Some systematic uncertainties are not easily implemented by varying model parameters. These are the subjects of ``simulated data" studies, where simulated data generated from a variant model is analyzed under the assumptions of the default model. Studies include 
varying 
$E_{\mathrm{b}}$, 
replacing the RFG model with a local Fermi gas model~\cite{nieves} or a spectral function model~\cite{Benhar-SF}, changing the 2p-2h model to an alternate one~\cite{Martini-2p2h} or fixing the 2p-2h model to be fully ``$\Delta$-like" or ``non-$\Delta$-like", varying the axial nucleon form factor to allow more realistic high $Q^2$ uncertainties~\cite{z_expansion,twoComponentAxial},
 and using an alternative single pion production model described in \cite{Kabirnezhad:2017jmf}. Additional simulated data studies, based on an excess observed at low muon momentum ($p_{\mu} \leq 400$ MeV) and moderate angle ($0.6 \leq \cos\theta_{\mu} \leq 0.8$) in the near detector, quantified possible biases in neutrino energy reconstruction by modeling this as an additional {\em ad hoc} interaction under hypotheses that it had 1p-1h, $\Delta$-like 2p-2h or non-$\Delta$-like 2p-2h kinematics. Finally, a discrepancy in the pion kinematic spectrum observed at the near detector motivated a simulated data study to check the impact on the signal samples at SK. 

Fits to these simulated data sets showed no significant biases in $\delta_{CP}$ or sin$^2\theta_{13}$; however biases in $\Delta m^2$ comparable to the total systematic uncertainty were seen for most data sets. This bias was accounted for by adding an additional source of uncertainty into the confidence intervals in $\Delta m^2$, as described later. As well as biases in $\Delta m^2$, fits to the varied $E_{\mathrm{b}}$ simulated data sets also showed biases in $\sin^2\theta_{23}$ comparable to the total systematic uncertainty. To account for this bias, an additional degree of freedom was added to the fit, which allows the model to replicate the spectra expected at the far detector when $E_{\mathrm{b}}$ is varied. After the addition of these additional uncertainties, fits to the simulated data sets no longer show biases that are significant compared to the total systematic error. 

{\it Fit to the near detector data. --- } Fitting the unoscillated spectra of CC candidate events in ND280 constrains the systematic parameters in the neutrino flux and cross-section models \cite{Abe:2011ks}. The CC samples are composed of reconstructed interactions in one of the two Fine-Grained Detectors (FGD) with particle tracking through Time Projection Chambers (TPC) interspersed among the FGDs. While both FGDs have active layers of segmented plastic scintillator, the second FGD (FGD2) additionally contains six water-target modules, allowing direct constraints of neutrino interactions on H$_2$O, the same target as SK. 
The ND280 event selection is unchanged from the previous T2K publication \cite{Abe:2017bay}. The CC inclusive events are separated into different samples depending on the FGD in which the interaction occurred, the beam mode, the muon charge and the final-state pion multiplicity.
The negative muon candidates from data taken in neutrino mode are divided into three samples per FGD based on reconstructed final state topologies: no pion candidate (CC0$\pi$), one $\pi^+$ candidate (CC1$\pi$), and all the other CC event candidates (CC Other), dominated respectively by the CCQE-like process, CC single pion production, and deep inelastic scattering. In antineutrino mode, positively- and negatively-charged muon tracks are used to define CC event candidates, which are distributed in two topologies: those with only a single muon track reconstructed in the TPC (CC 1-track), and those with at least one other track reconstructed in the TPC (CC N-track).
All event samples are binned according to the candidate's momentum $p_\mu$ and $\cos\theta_\mu$, where $\theta_\mu$ is the angle between the track direction and the detector axis. A binned likelihood fit to the data is performed assuming a Poisson-distributed number of events in each bin with an expectation computed from the flux, cross-section and ND280 detector models. The near detector systematic and flux parameters are marginalized in estimating the far detector flux and cross-section parameters and their covariances. The uncertainties on  neutral current and $\nu_e$ interactions cannot be constrained by the current ND280 selection, therefore the fit leaves the related parameters unconstrained.
\begin{figure}[h]
\centering
  \includegraphics[width=\columnwidth]{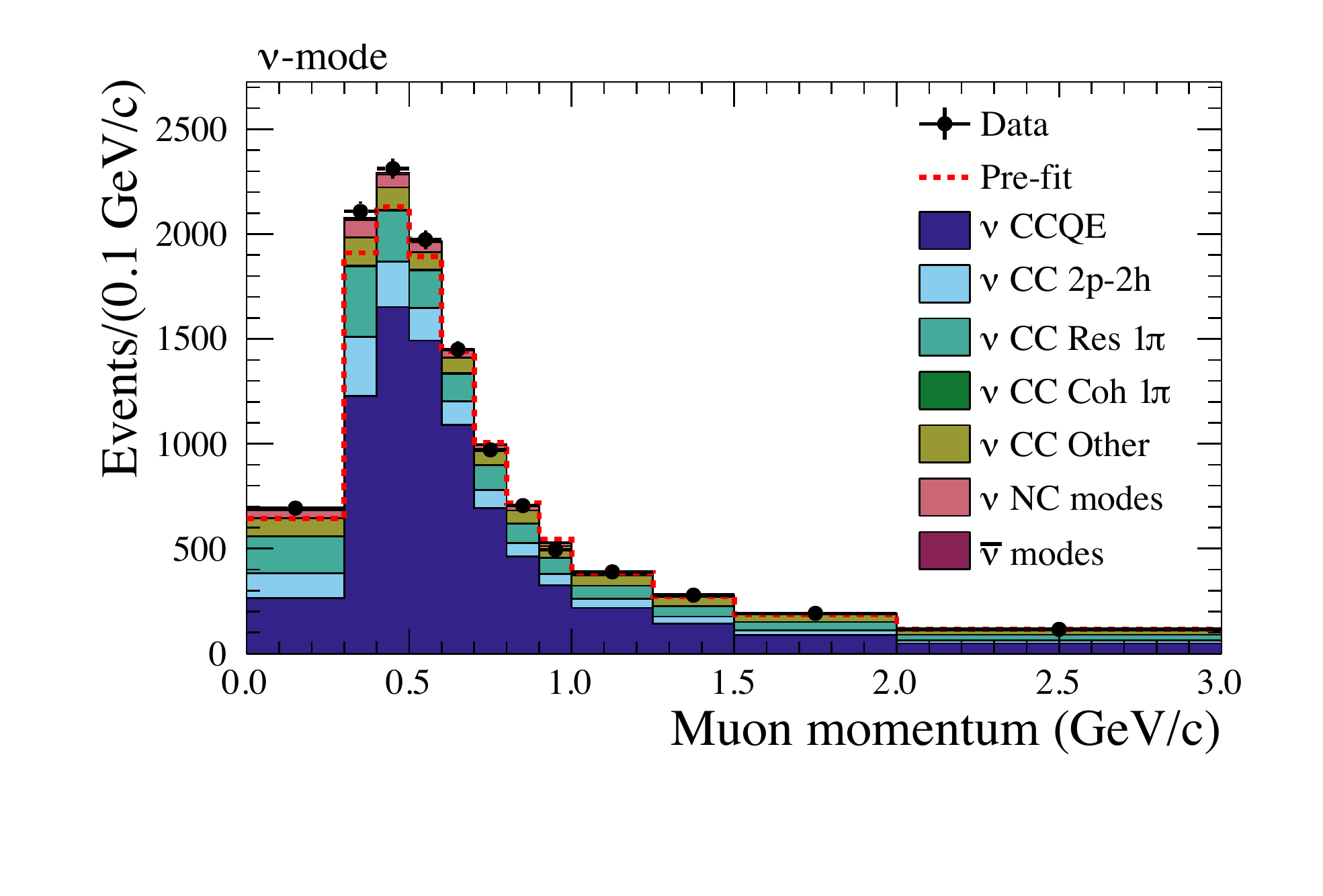} \\\vspace*{-5ex}
\caption{FGD2 data, and model predictions prior to and after ND280 data fit, binned in $p_\mu$ for the $\nu$ beam mode CC$0\pi$ sample. The prediction after the ND280 data fit is separated by type of interaction.}
\label{fig:banff}
\end{figure}
Figure~\ref{fig:banff} shows data, pre-fit and post-fit Monte-Carlo $p_\mu$ distributions for the FGD2 CC0$\pi$ sample. A deficit of $10$\%--$15\%$ in the pre-fit predicted number of events is observed, which is consistent with the previous T2K publications \cite{Abe:2017bay}. In this previous analysis, the simulated flux was increased to compensate the deficit. This is now resolved by the new RPA treatment, by increasing the low $Q^2$ part of the cross section.
Good agreement is observed between the post-fit model and the data, with a \emph{p}-value of 0.473, which is better agreement than in the previous T2K publication \cite{Abe:2017bay}, partly due to the modified cross-section parametrization.
The fit to the ND280 data reduces the flux and the ND280-constrained interaction model uncertainties on the predicted event rate at the far detector from 11--14\% to 2.5--4\% for the different samples.

\textit{Far detector event selection and data. --- }
Events at the far detector are required to be time-coincident with the beam and to be fully contained in the SK inner detector, by requiring limited activity in the outer detector. A newly-deployed Cherenkov-ring reconstruction algorithm, previously used only for neutral current (NC) $\pi^0$ background suppression~\cite{Abe:2015awa}, is used to classify events into five analysis samples, enriched in: $\nupbar_\mu$ CCQE; $\nupbar_e$ CCQE; and $\nu_e$ CC1$\pi^+$ where the $\pi^+$ is below Cherenkov threshold. The reconstruction algorithm uses all the information in an event by simultaneously fitting the time and charge of every photosensor in the detector. This results in an improved resolution of reconstructed quantities and particle identification. 

The fiducial volume is defined for each sample in terms of the minimum distance between the neutrino interaction vertex and the detector wall (\emph{wall}) and the distance from the vertex to the wall in the direction of propagation (\emph{towall}). These criteria are optimized taking into account both statistical and systematic uncertainties, with the systematic parameters related to ring-counting and $e$/$\mu$, $e$/$\pi^0$ and $\mu$/$\pi^+$ separation being constrained in a fit to SK atmospheric data. 
Other systematic uncertainties related to the modeling of the far detector are estimated using non-neutrino control samples.
Detector systematic error covariances between samples and bins for the oscillation analysis are constructed in the same way as was described in previous T2K publications~\cite{Abe:2015awa}.

The $\pi^0$ and $\pi^+$ NC suppression cuts are optimized by running a simplified oscillation analysis~\cite{Abe:2014tzr} on a simulated data set and choosing the criteria that minimize the uncertainty on the oscillation parameters. 

\begin{figure}[h]                                                                                                                                                                                                  \centering
    \includegraphics[width=\columnwidth]{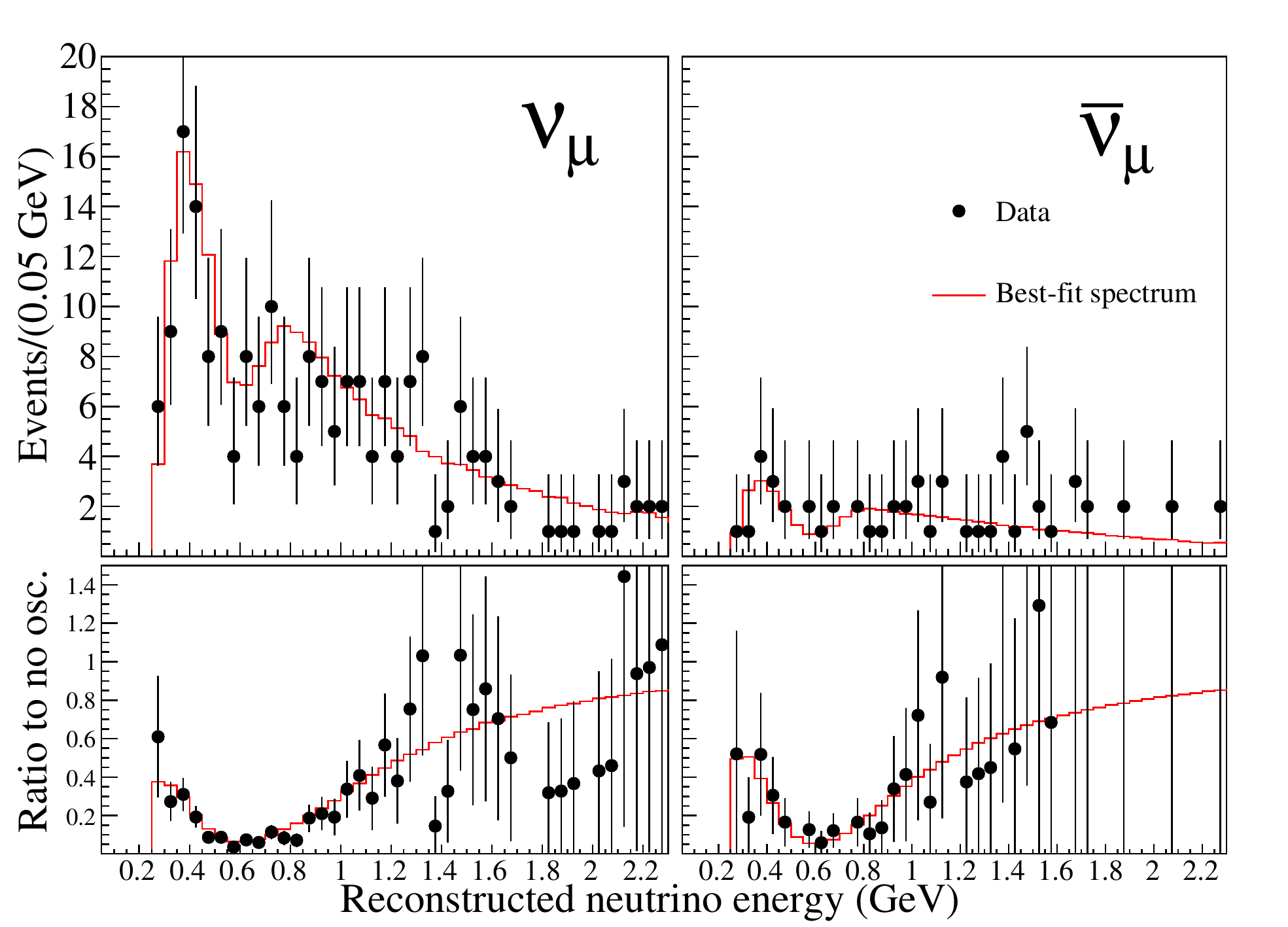}
    \caption{Reconstructed neutrino energy distributions at the far detector for the $\nu_\mu$ CCQE (left) and $\bar{\nu}_\mu$ CCQE (right) enriched samples with total predicted event rate shown in red. Ratios to the predictions under the no oscillation hypothesis are shown in the bottom figures.}
    \label{fig:erec_numu_plots}
\end{figure}

All selected events are required to have only one Cherenkov ring. For the $\nupbar_{\mu}$ CCQE-enriched samples, the single-ring events are further required to have $wall >$  50~cm and $towall >$ 250~cm; be classified as $\mu$-like by the $\mu$/$e$ separation cut; have a reconstructed momentum greater than 200~MeV/$c$; have up to one decay-electron candidate; and satisfy the $\pi^+$ rejection criterion. After these selection cuts are applied, 240 events are found in the neutrino-mode data and 68 in antineutrino-mode data, with an expectation of 261.6 and 62.0, respectively, for sin$^2 \theta_{23} = 0.528$ and $\Delta m_{32}^2 = 2.509 \times 10^{-3}$eV$^2$/$c^4$. The $E_\mathrm{rec}$ distributions for the data and best-fit Monte Carlo are shown in Figure~\ref{fig:erec_numu_plots}.

\begin{figure}[h]                                                                                                                                                                                                  \centering
    \includegraphics[width=\columnwidth]{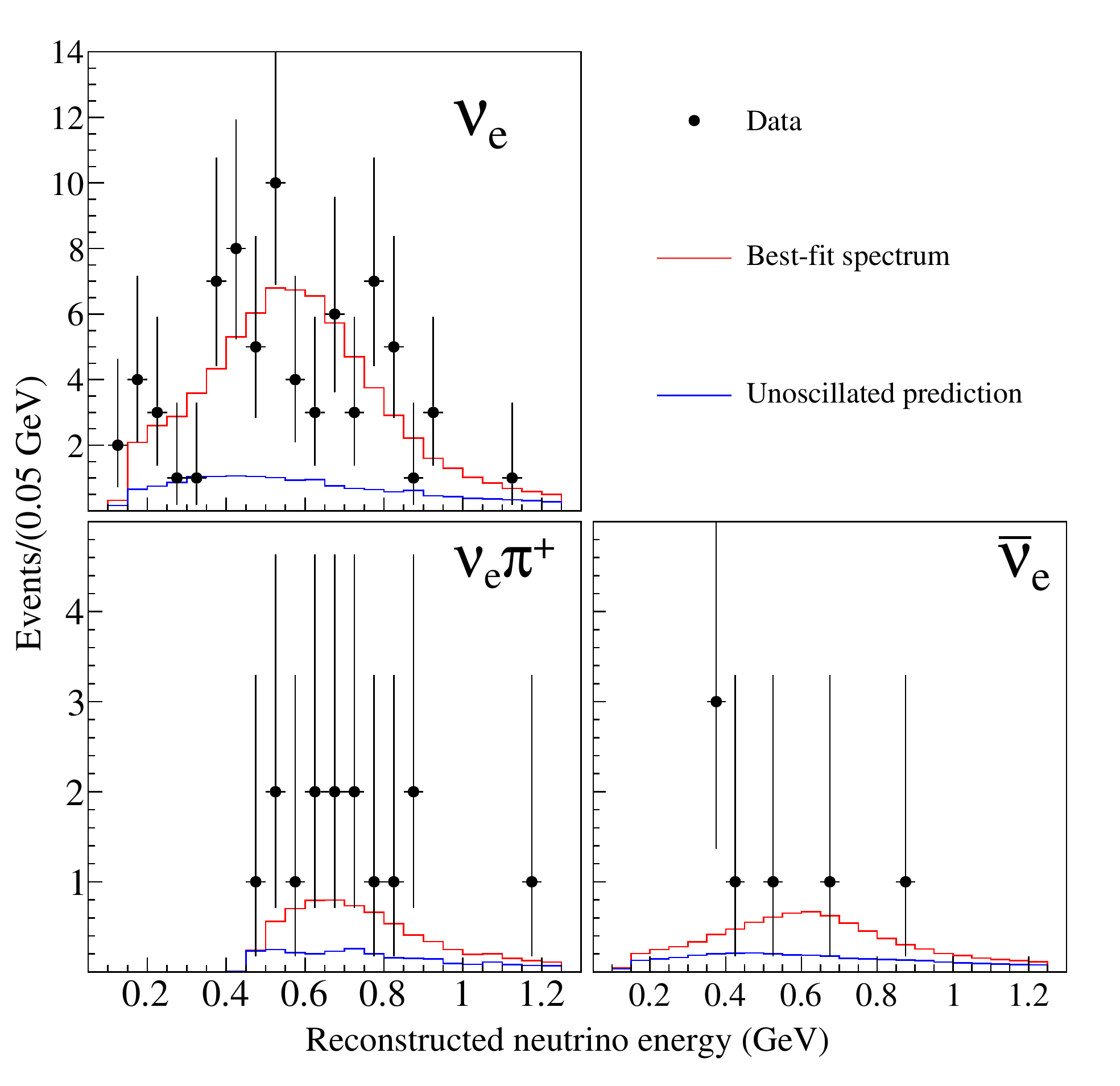}
    \caption{Reconstructed neutrino energy distributions at the far detector for the $\nu_e$ CCQE (top left), $\nu_e$ CC1$\pi^+$ (bottom left) and $\bar{\nu}_e$ CCQE (bottom right) enriched samples. Predictions under the no oscillation hypothesis are shown in blue and best-fit spectra in red.}
    \label{fig:erec_plots}
\end{figure}

The $\nupbar_{e}$ CCQE-enriched samples contain $e$-like events with no decay electron candidates, that pass the $\pi^0$ rejection cut, have $wall >$ 80~cm, $towall >$ 170~cm, momentum $>$ 100~MeV/$c$, and a reconstructed neutrino energy ($E_\mathrm{rec}$) lower than 1250~MeV. $E_{rec}$ is calculated from the lepton momentum and angle assuming CCQE kinematics. The $\nu_e$ CC1$\pi^+$-enriched sample has the same selection criteria with the exception of the fiducial volume criteria, which are $wall >$ 50~cm and $towall >$ 270~cm, and the requirement of one decay electron candidate in the event, from which the presence of a $\pi^+$ is inferred. Like in the case of the CCQE-enriched samples, $E_\mathrm{rec}$ for the $\nu_e$ CC1$\pi^+$ sample is calculated from the outgoing electron kinematics, except in this case the $\Delta^{++}$ mass is assumed for the outgoing nucleon.
Event yields for these samples are compared to Monte-Carlo predictions in Table~\ref{tbl:rates_dcp} and their $E_\mathrm{rec}$ distributions are shown in Figure~\ref{fig:erec_plots}.

Compared to previous T2K publications, the optimized event selection criteria are expected to increase the acceptance for
$\nupbar_{\mu}$ CCQE events by 15\% with a 50\% reduction of the NC1$\pi^+$ background; to increase the $\nupbar_{e}$ CC events acceptance by 20\% with similar purity to previous analyses; and to increase the $\nu_{e}$ CC1$\pi^+$ acceptance by 33\% with a 70\% reduction in background caused by particle misidentification. 
%
%
A summary of the systematic uncertainties on the predicted event rates at SK is given in Table~\ref{tbl:syst}.
%
\begin{table}[h]
\begin{center}
\caption{Systematic uncertainty on far detector event yields.}
\label{tbl:syst}
\begin{tabular}{l|c|c|c|c|c}
\hline \hline
Source [\%]                                & $\nu_\mu$ & $\nu_e$ & $\nu_e\pi^+$ & $\bar{\nu}_\mu$ & $\bar{\nu}_e$ \\ \hline
ND280-unconstrained cross section          & 2.4         & 7.8       &  4.1           &  1.7              &  4.8            \\
Flux \& ND280-constrained cross sec.      & 3.3       & 3.2     &  4.1        &  2.7            &  2.9          \\
SK detector systematics                    & 2.4       & 2.9     &  13.3        &  2.0            &  3.8          \\
Hadronic re-interactions     & 2.2       & 3.0     &  11.5        &  2.0            &  2.3          \\ \hline
Total                                      & 5.1       & 8.8     &  18.4        &  4.3            &  7.1          \\
\hline \hline
\end{tabular}
\end{center}
\end{table}

\textit{Oscillation analysis.}---A joint maximum-likelihood fit to five far-detector samples constrains the oscillation parameters sin$^2\theta_{23}$, $\Delta m^{2}$, sin$^2\theta_{13}$ and $\delta_{CP}$. Oscillation probabilities are calculated using the full three-flavor oscillation formulas~\cite{PhysRevD.22.2718} including matter effects, with a crust density of $\rho = 2.6~$g/cm$^3$~\cite{Hagiwara2011}.

Priors for the flux and interaction cross-section parameters are obtained using results from a fit to the near-detector data. Flat priors are chosen for sin$^2\theta_{23}$, $|\Delta m^{2}|$ and $\delta_{CP}$. The two mass orderings are each given a probability of 50\%. In some fits a flat prior is also chosen for sin$^2 2\theta_{13}$; whereas, in fits that use reactor neutrino measurements, we use a Gaussian prior of sin$^2 2\theta_{13}=0.0857 \pm 0.0046$~\cite{PDG}. The $\theta_{12}$ and $\Delta m^2_{21}$ parameters have negligible effects and are constrained by Gaussian priors from the PDG~\cite{PDG}.

Using the same procedure as~\cite{Abe:2017vifw}, we integrate the product of the likelihood and the nuisance priors to obtain the marginal likelihood, which does not depend on the nuisance parameters. We define the marginal likelihood ratio as $-2\Delta$ln$\mathcal{L} = -2$ ln$(\mathcal{L}/\mathcal{L}_{\rm{max}})$, where $\mathcal{L}_{\rm{max}}$ is the maximum marginal likelihood. 

Using this statistic, three independent analyses have been developed. The first and second analyses provide confidence intervals using a hybrid Bayesian-frequentist approach~\cite{COUSINS1992331}. The third analysis provides credible intervals using the posterior probability distributions calculated with a fully Bayesian Markov chain Monte Carlo method~\cite{MCMC}. This analysis also simultaneously fits both near- and far-detector data, which validates the extrapolation of nuisance parameters from the near to far detector. For all three analyses, the $\nupbar_{\mu}$ samples are binned by $E_{\mathrm{rec}}$. The first and third analyses bin the three $\nupbar_{e}$ samples in $E_{\mathrm{rec}}$ and lepton angle, $\theta$, relative to the beam, while the second analysis uses lepton momentum, $p$, and $\theta$. All three analyses give consistent results.

Expected event rates for various values of $\delta_{CP}$ and mass ordering are shown in Table~\ref{tbl:rates_dcp}. An indication of the sensitivity to $\delta_{CP}$ can be seen from the $\sim 20\%$ variation in the predicted total event rate between the $CP$ conserved case ($\delta_{CP}=0, \pi$) and when $CP$ is maximally violated. The $\nupbar_{\mu}$ event rates are negligibly affected by the mass ordering, whereas the $\nupbar_{e}$ rates differ by $\sim 10\%$ between mass orderings. In the $\nu_e$ CC1$\pi^+$ sample we see 15 events when we expected 6.9 for $\delta_{CP}=-\pi/2$ and normal ordering. The p-value to observe an upwards or downwards fluctuation of this significance in any one of the five samples used is 12\%.
The p-value to observe the data given the posterior expectation across all samples is greater than 35\%.

\begin{table}[h]
\begin{center}
\caption{ Number of events expected in the $\nu_e$ and $\overline{\nu}_e$ enriched samples for various values of $\delta_{CP}$ and both mass orderings compared to the observed numbers. The $\theta_{12}$ and $\Delta m^2_{21}$ parameters are assumed to be at the values in the PDG. The other oscillation parameters have been set to: $\sin^2\theta_{23}=0.528$, $\sin^2\theta_{13}=0.0219$, $|\Delta m^2| = 2.509\times 10^{-3} \rm{eV}^2c^{-4}$.}
\label{tbl:rates_dcp}

\begin{tabular}{ll|ccc}
\hline 
\hline
 & $\delta_{CP}$ & $\nu_e$ CCQE & $\nu_e$CC 1$\pi^+$ & $\overline{\nu}_e$ CCQE \\
\hline
         & $-\pi/2$     & 73.5 & 6.9 & 7.9  \\
Normal   & $0$          & 61.4 & 6.0 & 9.0  \\
ordering & $\pi/2$      & 49.9 & 4.9 & 10.0 \\
         & $\pi$        & 61.9 & 5.8 & 8.9  \\
\hline
         & $-\pi/2$     & 64.9 & 6.2 & 8.5  \\
Inverted & $0$          & 54.4 & 5.1 & 9.8  \\
ordering & $\pi/2$      & 43.5 & 4.3 & 10.9 \\
         & $\pi$        & 54.0 & 5.3 & 9.7  \\
\hline
Observed &              & 74   & 15  & 7    \\
\hline
\hline
\end{tabular}
\end{center}
\end{table}

\begin{figure}[h]
  \centering
    \includegraphics[width=\columnwidth]{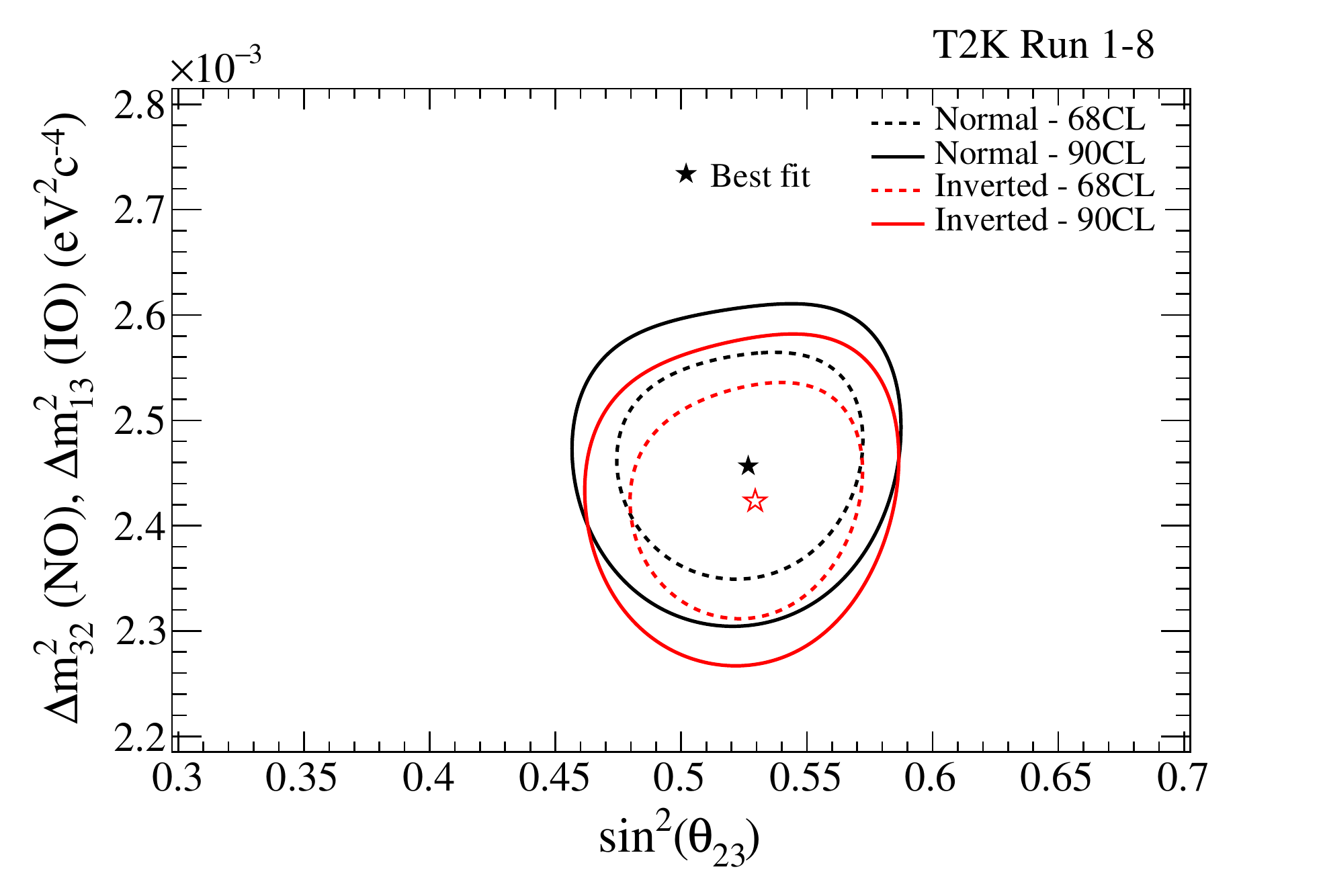}
    \caption{The 68\% (90\%) constant $-2\Delta$ln$\mathcal{L}$ confidence regions in the $|\Delta m^{2}|$-sin$^{2}\theta_{23}$ plane for normal (black) and inverted (red) ordering using the reactor measurement prior on sin$^{2}(2\theta_{13})$.}
    \label{fig:oa_dis}
\end{figure}
\begin{figure}[h]
  \centering
    \includegraphics[width=\columnwidth]{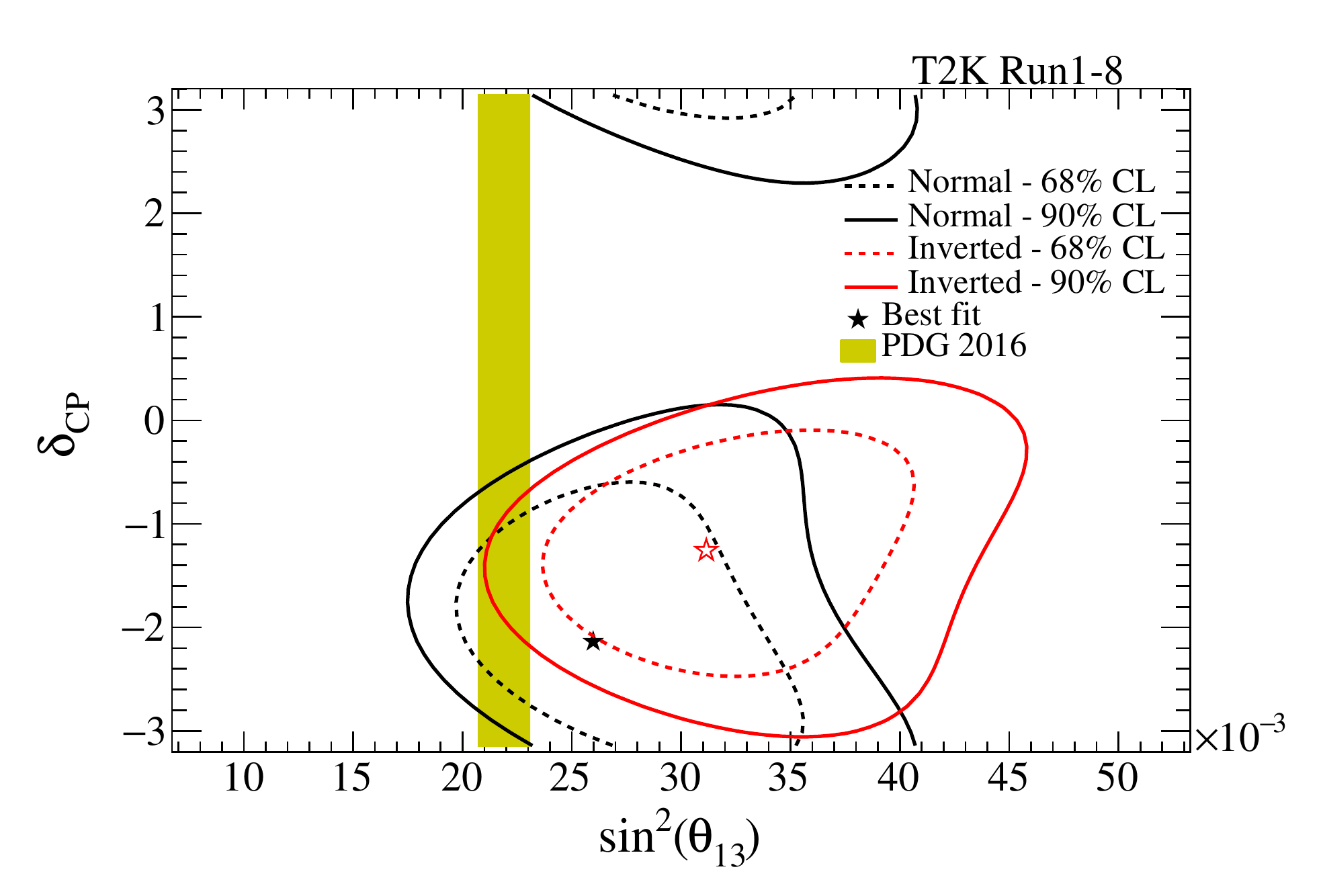}
    \caption{The 68\% (90\%) constant $-2\Delta$ln$\mathcal{L}$ confidence regions in the sin$^2\theta_{13}$-$\delta_{CP}$ plane using a flat prior on sin$^2(2\theta_{13})$, assuming normal (black) and inverted (red) mass ordering. The 68\% confidence region from reactor experiments on sin$^2\theta_{13}$ is shown by the yellow vertical band.}
    \label{fig:oa_app}
\end{figure}
\begin{figure}[h]
  \centering
    \includegraphics[width=\columnwidth]{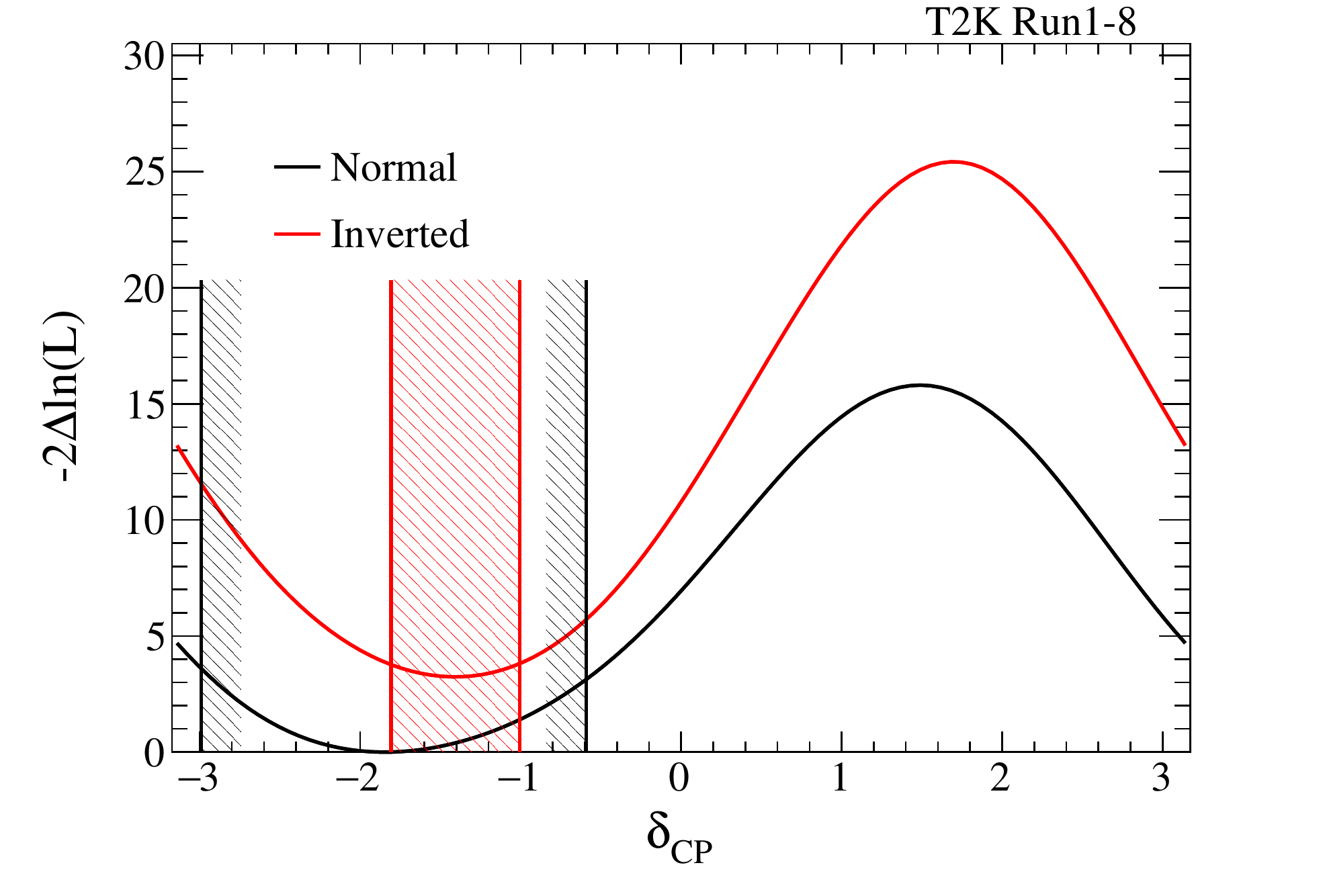}
    \caption{1D $-2\Delta$ln$\mathcal{L}$ as a function of $\delta_{CP}$ for normal (black) and inverted (red) mass ordering using the reactor measurement prior on sin$^{2}(2\theta_{13})$. The vertical lines show the corresponding allowed 2$\sigma$ confidence intervals, calculated using the Feldman-Cousins method instead of the constant $-2\Delta$ ln$\mathcal{L}$ method.}
    \label{fig:oa_dcp}
\end{figure}

Fits to determine either one or two of the oscillation parameters are performed, while the other parameters are marginalized. The constant $-2\Delta $ln$\mathcal{L}$ method is then used to set confidence regions~\cite{PDG}. Confidence regions in the $|\Delta m^{2}|$-sin$^{2}\theta_{23}$ plane (Fig.~\ref{fig:oa_dis}) were first computed for each mass ordering separately using the reactor measurement prior on sin$^{2} \theta_{13}$. The likelihood used to generate these confidence regions is convolved with a Gaussian function in the $\Delta m^2$ direction. The standard deviation of this Gaussian is $3.5\times 10^{-5}$~eV$^{2}/c^{4}$, which is the quadrature sum of the biases on $\Delta m^2$ seen in the fits to the simulated data sets. 

The best-fit values and the $1\sigma$ errors of $\sin^2\theta_{23}$ and $\Delta m^2$ are $0.526^{+0.032}_{-0.036}$ $(0.530^{+0.030}_{-0.034})$ and $2.463^{+0.071}_{-0.070}\times 10^{-3}(2.432\pm 0.070\times10^{-3})$ $ \mathrm{eV}^2/c^4$ respectively for normal (inverted) ordering. The result is consistent with maximal disappearance and the posterior probability for $\theta_{23}$ to be in the second octant (sin$^2\theta_{23}>0.5$) is 78\%.
The $\Delta m^2$ value is consistent with the Daya Bay reactor measurement~\cite{PhysRevD.95.072006}. 

Confidence regions in the sin$^2\theta_{13}$-$\delta_{CP}$ plane were calculated, without using the reactor measurement prior on sin$^{2}(2\theta_{13})$, for both the normal and inverted orderings (Fig.~\ref{fig:oa_app}). T2K's measurement of sin$^2\theta_{13}$ agrees well with the reactor measurement. 

Confidence intervals for $\delta_{CP}$ were calculated using the Feldman--Cousins method~\cite{FC}, marginalized over both mass orderings simultaneously, from a fit using the reactor measurement prior. The best fit value is $\delta_{CP}=-1.87 (-1.43)$ for the normal (inverted) ordering, which is close to maximal $CP$ violation (Fig.~\ref{fig:oa_dcp}). The $\delta_{CP}$ confidence intervals at $2\sigma$ (95.45\%) are ($-2.99$, $-0.59$) for normal ordering and ($-1.81$, $-1.01$) for inverted ordering. Both intervals exclude the $CP$-conserving values of 0 and $\pi$. The Bayesian credible interval at 95.45\% is ($-3.02$,$-0.44$), marginalizing over the mass ordering. The normal ordering is preferred with a posterior probability of 87\%.

Sensitivity studies show that, if the true value of $\delta_{CP}$ is $-\pi/2$ and the mass ordering is normal, 22\% of simulated experiments exclude $\delta_{CP}=0$ and $\pi$ at $2\sigma$ C.L.

\textit{Conclusions. --- }
T2K has constrained 
the leptonic $CP$ violation phase ($\delta_{CP}$), $\sin^2\theta_{23}$, $\Delta m^2$ and the posterior probability for the mass orderings
with additional data and with an improved event selection efficiency.
The 2$\sigma$ (95.45\%) confidence interval for $\delta_{CP}$ does not contain the $CP$-conserving values of $\delta_{CP}=0,\pi$ for either of the mass orderings.
The current result is predominantly limited by statistics.
T2K will accumulate 2.5 times more data, thereby improving sensitivity for the relevant oscillation parameters.
The data related to the measurement and results presented in this Letter can be found in \cite{data}.

\ifnum\sizecheck=0
\begin{acknowledgments}

We thank the J-PARC staff for superb accelerator performance. We thank the
CERN NA61/SHINE Collaboration for providing valuable particle production data.
We acknowledge the support of MEXT, Japan;
NSERC (Grant No. SAPPJ-2014-00031), NRC and CFI, Canada;
CEA and CNRS/IN2P3, France;
DFG, Germany;
INFN, Italy;
National Science Centre (NCN) and Ministry of Science and Higher Education, Poland;
RSF, RFBR, and MES, Russia;
MINECO and ERDF funds, Spain;
SNSF and SERI, Switzerland;
STFC, UK; and
DOE, USA.
We also thank CERN for the UA1/NOMAD magnet,
DESY for the HERA-B magnet mover system,
NII for SINET4,
the WestGrid, SciNet and CalculQuebec consortia in Compute Canada,
and GridPP and the Emerald High Performance Computing facility in the United Kingdom.
In addition, participation of individual researchers and institutions has been further
supported by funds from ERC (FP7), H2020 Grant No. RISE-GA644294-JENNIFER, EU;
JSPS, Japan;
Royal Society, UK;
the Alfred P. Sloan Foundation and the DOE Early Career program, USA.
  
\end{acknowledgments}
  \bibliography{2017OA}
\fi

\ifnum\PRLsupp=0
  \clearpage
appendix
\fi

\end{document}